
\magnification\magstep1
\scrollmode
\hfuzz=1pt


\font\tenbbb=msbm10 \font\sevenbbb=msbm7
\newfam\bbbfam             
\textfont\bbbfam=\tenbbb \scriptfont\bbbfam=\sevenbbb
\def\Bbb{\fam\bbbfam}      

\font\tengoth=eufm10 \font\sevengoth=eufm7
\newfam\gothfam            
\textfont\gothfam=\tengoth \scriptfont\gothfam=\sevengoth

\font\tenbm=cmmib10 \font\sevenbm=cmmib10 at 7pt
\newfam\bmfam              
\textfont\bmfam=\tenbm \scriptfont\bmfam=\sevenbm

\def\opname#1{\mathop{\rm#1}\nolimits} 

\def\Al{{\cal A}}                 
\def\Bl{{\cal B}}                 
\def\C{{\Bbb C}}                  
\def\cite#1{{\rm\lbrack#1\rbrack}}
\def\Cl{{\cal C}}                 
\def\eq#1{{\rm(#1)}}              
\def\H{{\cal H}}                  
\def\Ha{{\bar{\cal H}}}           
\def\HH{{\Bbb H}}                 
\def\ox{\otimes}                  
\def\Str{\opname{Str}}            
\def\stroke{\mathbin\vert}        
\def\tfrac#1#2{{\textstyle{#1\over#2}}} 
\def\Tr{\opname{Tr}}              
\def\U{{\cal U}}                  
\def\x{\times}                    
\def\7{\dagger}                   
\def\8{\bullet}                   
\def\.{\cdot}                     
\def\:{\colon}                    
\def\<#1,#2>{\langle#1\stroke#2\rangle} 

\outer\def\section#1. #2\par{
      \bigskip\bigskip
      \message{#1. #2}%
      \leftline{\bf#1. #2}\nobreak
      \smallskip\noindent}

\outer\def\subsection#1. #2\par{
      \bigskip \vskip\parskip
      \message{\string\S\space#1.}%
      \leftline{\it#1. #2}\nobreak
      \smallskip\noindent}

\def\declare#1. #2\par{\medskip   
      \noindent{\bf#1.}\rm
      \enspace\ignorespaces
      #2\par\smallskip}

\def\refno#1. #2\par{\smallskip   
      \item{\rm\lbrack#1\rbrack}
      #2\par}

\def\topic#1.#2. {\medskip\noindent  
          {\bf#1.#2.}\enspace\ignorespaces}


\def\Sirius{1}
\def\Book{2}
\def\KStris{3}
\def\Tresttalk{4}
\def\Proteus{5}
\def\Connesprivatecomm{6}
\def\lag{7}
\def\Ibanez{8}
\def\Orpheus{9}
\def\Persephone{10}
\rightline{FT/UCM/7/95}
\rightline{FT-UAM-95100}
\rightline{hep-th/9506115}
\bigskip\bigskip

\centerline{\bf Anomaly cancellation and the gauge group of the standard model
in NCG}
\bigskip
\centerline{\rm Enrique Alvarez \footnote*{and Universidad Aut\'onoma,
 28049 Madrid, Spain}}
\medskip
\centerline{ CERN-TH, 1211 Geneva 23, Switzerland}
\bigskip
\centerline{\rm Jos\'e M. Gracia-Bond\'{\i}a}
\medskip
\centerline{Escuela de Matem\'atica, Universidad de Costa Rica,  2060
 San
Jos\'e, Costa Rica}
\bigskip

\centerline{\rm C.P. Mart\'{\i}n}
\medskip

\centerline{ Departamento de F\'{\i}sica Te\'orica I,
Universidad Complutense, 28040 Madrid, Spain}

\bigskip\bigskip


\begingroup\narrower\parindent=1em
It is well known that anomaly cancellation {\it almost} determines the
 hypercharges in the standard model. A related (and somewhat more stronger)
 phenomenon takes
place in Connes' NCG framework: unimodularity (a technical condition on
elements
of the algebra) is {\it strictly} equivalent to anomaly cancellation (in
the absence of right-handed neutrinos); and this in turn reduces the symmetry
group of the theory to the standard $SU(3) \times SU(2) \times U(1)$.
\par\endgroup 

\section 1. Introduction

There is a deep relationship between anomaly cancellation and the actual
values of the hypercharges in the standard model; it is well known [\Ibanez]
that anomaly cancellation only allows for two solutions: the ``physical''
one, and a ``bizarre'' solution, with all the hypercharges zero except for
the $\bar u$ and $\bar d$, whose sum must vanish.

One of the many fascinating aspects of the Connes(-Lott) approach to the
standard model through Non-Commutative Geometry (NCG) is that the
demand of anomaly cancellation is fulfilled through a mathematical restriction
on the elements of the algebra, technically called {\it unimodularity}
(somewhat similar to the restriction to unit determinant elements in a
unitary group).

It is remarkable that a quite subtle {\it quantum} property, such as
anomaly cancellation, is achieved automatically after imposing an
apparently unrelated and much simpler condition, the unimodularity
condition. Our purpose in this paper is to explore
this relationship.
After a short reminder of the NCG set-up, we shall explore the converse
property, namely, to what extent is true that anomaly cancellation does
imply unimodularity.
We shall find that, under certain conditions, {\it they are strictly
equivalent}
so that one can say, slightly overstating
it, that anomaly cancellation, in the NCG context, determines the gauge
group in the representation observed in Nature.

\section 2. Connes' unimodularity conditions

The ``old scheme" for the NCG reconstruction of the SM has been spelled out
in [\Sirius, \Book, \KStris].  Recently Connes introduced a ``new scheme"
[\Tresttalk].  The key element of both is an associative algebra (the
noncommutative spacetime) represented by operators on the Hilbert space
$\H \oplus \Ha$ of all fermions.  The noncommutative gauge potential and gauge
field are forms on the noncommutative space, defined via successive
commutation with a Dirac-Yukawa operator.  We need to consider only one
quark and lepton family at a time.  Thus, when no right handed
neutrinos are present:
$$
\eqalignno{
&\H = \H_\ell \oplus \H_q
\cr
&:=    L^2(S) \ox \pmatrix{\C_{e;R} \cr \C^2_{e,\nu;L} \cr}
\,\oplus\, L^2(S) \ox \pmatrix{(\C_{d;R} \oplus \C_{u;R}) \ox \C^{N_c} \cr
       \C^2_{d,u;L} \ox \C^{N_c} \cr}
& (2.1) \cr}
$$
where $S$ denotes the space of spinors, $N_c$ the color degrees of freedom;
and similarly for the conjugate space $\bar\H = \bar\H_\ell \oplus \bar\H_q$ of
antiparticles. When right handed neutrinos are present, $\H_\ell$ in
\eq{2.1} is replaced by
$$
L^2(S) \ox \pmatrix{\C_{e;R} \oplus \C_{\nu;R}
\cr \C^2_{e,\nu;L}}.
$$

In the old scheme there is actually a pair of algebras $(\Al, \Bl)$,
with compatible actions on $\H \oplus\Ha$. They are the tensor product of the
commutative algebra of smooth functions over the ordinary spacetime by the
finite-part algebras
$$
\Al_F := \C \oplus \HH,  \qquad  \Bl_F := \C \oplus M_3(\C),
$$
representing respectively the flavour and colour degrees of freedom; here
$\HH$ is the algebra of quaternions of Hamilton. In the new scheme they are
replaced by a single algebra $\Cl$ with finite part $\Cl_F := \C \oplus \HH
\oplus M_3(\C)$.

Denote by $\Psi$ a generic element of the fermion space. The gauge invariant
action associated to the fermion fields
$$
I (\Psi) = \<\Psi, (D + A_{NC})\Psi>
\eqno (2.2)
$$
gives rise both to the kinetic and the Yukawa terms in the SM Lagrangian.
By construction, the old scheme theory is invariant under the direct product of
the groups of unitaries of $\Al$ and $\Bl$, which is $C^\infty(M, U(1) \x
SU(2)_L \x U(1) \x U(3))$. Thus, the noncommutative philosophy faces the
problem of finding a credible and useful way to reduce this group to a
$C^\infty(M, SU(2)_L \x U(1)_Y \x SU(3))$ subgroup. Note that we are not
allowed to change the representation of these groups, which is given by the
representation of the algebras.

Actually, $A_{NC}$ is the sum of the flavour and colour gauge potentials $A_f$
and $A_c$. Let $A,V,A',K$ be skewhermitian 1-forms with values in $\C$,
$\HH$, $\C$ and $M_3(\C)$ respectively. One has, for each fermionic family
with a right handed neutrino:
$$
A_f + A_c = \bordermatrix{&\scriptstyle {eR} &\scriptstyle {\nu R}
&\scriptstyle {dR} &\scriptstyle {uR} &\scriptstyle {\ell L}
&\scriptstyle {qL}
\cr
\scriptstyle {eR}      & A + A' & & & & & \cr
\scriptstyle {\nu R}   & & -A + A' & & & & \cr
\scriptstyle {dR}      & & & A + K & & & \cr
\scriptstyle {uR}      & & & & -A + K & & \cr
\scriptstyle {\ell L}  & & & & & V + A' & \cr
\scriptstyle {qL}      & & & & & & V + K \cr},
\eqno(2.3)
$$
plus a conjugate copy ($A \to -A$, $A' \to -A'$, $K \to -K$) in the
antiparticle space. We are forgetting about the off-diagonal terms, which are
unimportant here. If there are no right handed neutrinos,
just suppress the second row and column.

Connes finds in [\Book] a reduction rule, called the unimodularity condition on
the $\U(\Al) \x \U(\Bl)$ group, which can be rewritten [\Proteus] as
$$
\Tr_{\H_R}(A_f + A_c) = 0,  \qquad  \Tr_{\H_L}(A_f + A_c) = 0.
\eqno(2.4)
$$
Here $\H_R$ denote the space of the right-handed particles and $\H_L$
the space of the left-handed particles (the same conditions apply with the
same result on the antiparticle side). Now $V^* = -V$ means that
$V$ is a zero-trace quaternion, so $\Tr_{\H_L}(A_f) = 0$ automatically; thus
$\Tr_{\H_L}(A_c) = 0$, which yields the condition
$$
A' = - \Tr K.
$$
Let $N_1$ be the number of massive neutrino species. Then,
$$
\Tr_{\H_R}(A_f + A_c)
= (N_F + N_1) A' + (N_F - N_1) A + 2 N_F \Tr K.
$$
Combining both conditions, we get the reduction rule:
$$
A = A' = -\Tr K.
\eqno(2.5)
$$
as long as $N_1 < N_F$. On the other hand, if all species of neutrinos have
right handed components, the abelian part $A$ of the flavour gauge potential
remains free [\Proteus]. This is to be regarded as a drawback of the old
scheme.

When condition \eq{2.5} holds, we can identify $A = A' = -\Tr K$ in \eq{2.3} as
the generator of the $U(1)_Y$ physical gauge field. Thus the abelian part
of the noncommutative gauge potential reads:
$$
\bordermatrix{&\scriptstyle {eR} &\scriptstyle {\nu R}
&\scriptstyle {dR} &\scriptstyle {uR} &\scriptstyle {\ell L}
&\scriptstyle {qL}
\cr
\scriptstyle {eR}      & 2A & & & & & \cr
\scriptstyle {\nu R}   & & 0 & & & & \cr
\scriptstyle {dR}      & & & \tfrac23 A & & & \cr
\scriptstyle {uR}      & & & & -\tfrac43 A & & \cr
\scriptstyle {\ell L}  & & & & & A & \cr
\scriptstyle {qL}      & & & & & & -\tfrac13 A\cr}
\oplus\,
\bordermatrix{&\scriptstyle {\bar e L} &\scriptstyle {\bar \nu L}
&\scriptstyle {\bar d L} &\scriptstyle {\bar u L} &\scriptstyle {\bar\ell R}
&\scriptstyle {\bar q R}
\cr
\scriptstyle {\bar e L}      & -2A & & & & & \cr
\scriptstyle {\bar \nu L}   & & 0 & & & & \cr
\scriptstyle {\bar d L}      & & & -\tfrac23 A & & & \cr
\scriptstyle {\bar u L}      & & & & \tfrac43 A & & \cr
\scriptstyle {\bar\ell R}  & & & & & -A & \cr
\scriptstyle {\bar q R}      & & & & & & \tfrac13 A\cr}.
\eqno (2.6)
$$
In summary, we have killed two
extra $U(1)$ fields and we get in addition the tableau of hypercharge
assignments of the SM: if we conventionally adopt $Y(e_L, \nu_L) = -1$, there
follows for left-handed (anti-)leptons and (anti-)quarks: $Y(\bar e_L) = 2$;
$Y(\bar\nu_L) = 0$; $Y(\bar d_L) = \tfrac23$; $Y(\bar u_L) = -\tfrac43$ and
$Y(d_L, u_L) = \tfrac13$.

\smallskip

The matter of gauge invariance is more subtle in the new scheme. An important
role is played by the antilinear isometry $J$ that interchanges the
particle and antiparticle subspaces:
$$
J(\Psi, \bar\Xi) = (\Xi, \bar\Psi) \;{\rm for}\; (\Psi, \bar\Xi) \in \H \oplus
\Ha.
$$
The action of the gauge group is no longer simply given by the restriction of
the $\Cl$ action; rather it is of the form $(\Psi, \bar\Xi) \mapsto uJuJ (\Psi,
\bar\Xi)$, for $u$ belonging to $C^\infty(M, U(1) \x SU(2)_L \x U(3))$. This
translates into a noncommutative fermionic action of the form \eq{2.2}, where
now $A_{NC} = \widetilde A + J\widetilde AJ$ with
$$
\widetilde A
 = \bordermatrix{&\scriptstyle {eR} &\scriptstyle {\nu R}
&\scriptstyle {dR} &\scriptstyle {uR} &\scriptstyle {\ell L}
&\scriptstyle {qL}
\cr
\scriptstyle {eR}      & A & & & & & \cr
\scriptstyle {\nu R}   & & -A & & & & \cr
\scriptstyle {dR}      & & & A  & & & \cr
\scriptstyle {uR}      & & & & -A  & & \cr
\scriptstyle {\ell L}  & & & & & V  & \cr
\scriptstyle {qL}      & & & & & & V \cr}
\oplus\,
\bordermatrix{&\scriptstyle {\bar e L} &\scriptstyle {\bar\nu L}
&\scriptstyle {\bar d L} &\scriptstyle {\bar u L} &\scriptstyle {\bar\ell R}
&\scriptstyle {\bar q R}
\cr
\scriptstyle {\bar e L}      & -A  & & & & & \cr
\scriptstyle {\bar\nu L}     & & -A  & & & & \cr
\scriptstyle {\bar d L}      & & & -K & & & \cr
\scriptstyle {\bar u L}      & & & & -K & & \cr
\scriptstyle {\bar\ell R}    & & & & & -A & \cr
\scriptstyle {\bar q R}      & & & & & & -K \cr}.
\eqno (2.7)
$$
For a fermion family without right-handed neutrino, suppress the second
row and column.

Now, following Connes, instead of \eq{2.4} we impose the single unimodularity
condition
$$
\Str A_{NC} = 0.
\eqno (2.8)
$$
%
Reasoning as above, one
gets again the reduction to the SM gauge group and hypercharges. Note that this
happens irrespectively of whether right-handed neutrinos are present (it turns
 out
[\Connesprivatecomm] that the old $(\Al, \Bl)$ scheme with all neutrinos
with right-handed components does not obey Poincar\'e duality ---therefore
is not a noncommutative manifold in the strict sense--- whereas the new $\Cl$
scheme is Poincar\'e selfdual in any case).

\section 3. Unimodularity = cancellation of anomalies.

In this section we shall show that either one of the two
unimodularity conditions, eqs. (2.4)
and (2.8) is equivalent to anomaly cancellation in the Standard Model,
when obtained within the NCG framework.
We will assume that each family is anomaly free by itself, and we will
allow for the possibility of right-handed neutrinos towards the end of the
 section.

 It is a fact that the unimodularity conditions [\Book], eqs (2.4) and (2.8)
on the NCG potentials
do select the representations of $ SU(3) \times SU(2) \times
U(1)_{Y}$ carried out by leptons and quarks as observed in Nature
which in obvious notation is, for the $\bar{ e}_L$,$l_L = (e_L,\nu_L)$,
$\bar {d}_L $, $\bar u_L$ and $ q_L = (d_L,u_L)$, respectively:

$$ (1,1,2) \oplus (1,2,-1) \oplus (\bar 3,1, 2/3) \oplus (\bar 3,1, - 4/3)
\oplus (3,2,1/3)
\eqno (3.1)
$$
It is then plain that unimodularity implies absence of anomalies.
Our aim in this section is to show that in a certain sense, the
reverse is true as well.
Let us note, first of all, that locally, $U(3) \sim SU(3) \times U(1)_K$.
The index $K$ in the abelian factor stresses the fact that it comes from
the traceful generator in $U(3)$. Now, prior to imposing any
unimodularity
condition, the NCG formalism yields a model with a gauge symmetry which is
either  $ G \sim SU(3) \times SU(2) \times U(1)_A \times U(1)_{A^{\prime}}
\times
U(1)_K$ (in the old scheme), or else
$G^{\prime}\sim SU(3) \times SU(2) \times U(1)_A \times U(1)_K$ (in the new
scheme).

Moreover, the representations are not arbitrary; rather, they are fixed
by the corresponding representation of the NCG algebra over the Hilbert space
of the fermions. Actually, from eq. (2.3) we learn that the fermions
$\bar {e}_L$,$l_L$,$\bar{d}_L$,$\bar{u}_L$ and $q_L$ transform under $G$
as $$ (1,1,y,y',0)\oplus (1,2,0,-y',0)\oplus (\bar 3,1,y,0,k) \oplus
(\bar 3,1,-y,0,k)\oplus(3,2,0,0,-k),
\eqno (3.2)
$$
Where $y$, $y^{\prime}$ and $k$ set the $U(1)$ charge scales and do not
vanish.
Similarly, the $G^{\prime}$ charges in the new scheme are:
$$(1,1,2 y,0)\oplus (1,2,-y,0)\oplus (\bar 3 ,1,y,k) \oplus(\bar 3,1,-y,k)
\oplus (3,2,0,-k)
\eqno (3.3)
$$

All the NCG reasoning up to now has been classical. If from now on, one
 considers
the lagrangian so obtained as a standard quantum field theory, it is obvious
that one should impose cancellation of anomalies in order to get a sensible
theory. It is easy to see, though [\lag] that this theory is always
anomalous, due to the fact that, for $G$,
$$
tr Y_{A}^{3} =y^3
\eqno (3.4)
$$
and the $U(1)[SU(2)]^2$ anomalies, for $G^{\prime}$, are given by:
$$trY_A \{ T^{a}_{SU(2)},T^{b}_{SU(2)}\} = -2y
\eqno (3.5)
$$
(Both quantities are always different from zero).
This means that the group of unitaries of the algebra(s) does {\it not}
qualify as a consistent symmetry group at the quantum level
(and this is true in both the old and the new formulations).
A natural question to ask at this stage is whether there are subgroups
of $G$ or $G^{\prime}$ that are anomaly free (with the representation content
induced from the embedding).
We shall actually deal with a more modest version of the preceding, namely we
 shall
assume that the subgroups are of the form $SU(3) \times SU(2) \times H$, with
the same quantum numbers for $SU(3) \times SU(2)$ as indicated in
eqs. (3.1), (3.2) and (3.3).
Although those are certainly the simplest possibilities, they do not exhaust
them all. It is a quite natural restriction from the NCG point of
view, though,
because at any rate both the color and weak isospin structure are imposed by
 hand precisely for them to coincide with the standard model
and it does not seem wise to tamper with them more than necessary.
This leaves room for only two possibilities for $H$: either $U(1)$ or else
$U(1) \times U(1)$ (in the new scheme there is only the former possibility).

To be specific, what we are going to prove is that the {\it unique} subgroup
of $G$ ($G^{\prime}$), under the restrictions just specified, which is anomaly
free, is precisely, the standard model group , with the physical
representation
content. The actual embedding is defined, in the old scheme, through an
identification of $A$, $A^{\prime}$ and $K$, according to (2.5); and in the
new scheme, owing to the identification of $A$ and $K$ conveyed by (2.8).

In the old scheme there are two possibilities for $H$, namely $U(1) \times
U(1)$
and $U(1)$. There are three different ways of getting the first possibility, to
 wit
(representing the two abelian gauge fields of $H$ by $B$ and $B^{\prime}$
, and denoting by $\hat K =:{tr K \over 3}$):
$$
\eqalign{
i) A^{\prime}& = \alpha A + \beta \hat K
\cr
B &= A
\cr
B^{\prime}& = K
\cr
ii)\hat K& = \gamma A
\cr
B &= A
\cr
B^{\prime} &= A^{\prime}
\cr
iii)
A &= 0
\cr
B &= \hat K
\cr
B^{\prime} &= A^{\prime}
\cr}
\eqno (3.6)
$$
The linear constraints (3.6) now determine the $H$ quantum numbers
in terms of two arbitrary real parameters, $x$ and $y$, namely
$$
i)((1+\alpha)x,\beta y)\oplus(-\alpha x,-\beta y)\oplus (x,y) \oplus (-x,y)
\oplus(0,-y),
\eqno (3.7)
$$
\smallskip
$$
ii){(x,y) \oplus (0,-y) \oplus ((1+\gamma)x,0)\oplus((\gamma -1)x,0)\oplus
(-\gamma
x,0)},
\eqno (3.8)
$$
\smallskip
$$
iii){(0,y)\oplus(0,-y)\oplus(x,0)\oplus(x,0)\oplus(-x,0)},
\eqno (3.9)
$$

It is now a simple exercise to show that (3.7) is anomalous, since
$$
Tr Y_B^3 = ((1+\alpha)^3 - 2\alpha^3) x^3$$
$$
Tr Y_B \{ T^{a}_{SU(2)},T^{b}_{SU(2)}\} = - 2\alpha x,
\eqno (3.10)
$$
 and both expressions cannot vanish simultaneously, since $x \neq 0$.
The same thing happens with the representation conveyed by (3.8):
$$
Tr Y_{B^{\prime}}^3 = - y^3
$$
(always non-zero) as well as with the one coming from (3.9):
$$
Tr Y_{B^{\prime}}^3 = -y^3
$$
(again, never zero).This means that (3.6) never leads to an anomaly free
representation.

Let us now examine the other option, $H=U(1)$. The three different
possibilities for getting $H = U(1)$ are now (representing by $B$ the
only remaining abelian gauge field):
$$
\eqalign{
i) A &=\alpha A^{\prime}
\cr
\hat K &=\beta A^{\prime}
\cr
B&=A^{\prime}
\cr
ii)A&=\alpha \hat K
\cr
A^{\prime}&=0
\cr
B &=\hat K
\cr
iii)A^{\prime}&=0
\cr
\hat K &= 0
\cr
B &= A},
\eqno (3.11)
$$
The charges are then given in terms of a single parameter, $x\neq 0$:
$$
i)(1+\alpha)x\oplus-x\oplus (\alpha +\beta)x \oplus (\beta -\alpha)x
\oplus -\beta x,
\eqno (3.12)
$$
\smallskip
$$
ii)\alpha x \oplus 0\oplus(1+\alpha) x\oplus (\alpha -1)x\oplus -x,
\eqno (3.13)
$$
\smallskip
$$
iii)x\oplus 0\oplus x\oplus - x\oplus 0 ,
\eqno (3.14)
$$
Now it is easily checked that (3.13) and (3.14) are both anomalous, since,
for example, (3.13) implies:
$Tr Y_B \{ T^{a}_{SU(2)},T^{b}_{SU(2)} \} = -6x \neq 0$, and (3.14)
 leads in turn to:
$Tr Y_B^3 = x^3 \neq 0$. On the other hand, for (3.12) one gets
$$
Tr Y_B^3 = ((1+\alpha)^3 - 2 + 18 \alpha^2\beta) x^3$$
$$
Tr Y_B \{ T^{a}_{SU(2)},T^{b}_{SU(2)} \} = 2 (1+3 \beta)x,
\eqno (3.15)
$$
This means that (3.12) is anomaly-free if and only if
$$\alpha = 1 ;\;\;
\beta = -1/3,
\eqno (3.16)$$
\smallskip
This correspond to the fermionic representation of
$SU(3)\times SU(2)\times U(1)$:
\smallskip
$$
(1,1,2x)\oplus (1,2,-x)\oplus(\bar 3,1,2x/3 )\oplus(\bar 3,1,-4x/3)
 \oplus(3,2,x/3)$$
which coincides with the standard representation, up to normalization. Note
that if we substitute $\alpha = 1$ and $\beta = -1/3$ back in eq. (3.11)
one obtains $ A = A^{\prime} = - tr K $, so that one recovers the
unimodularity constraints given in eq (2.5).
\par
In the new scheme, the group of unitaries is $G^{\prime}$, with representation
content given by (3.3).With a by now familiar reasoning, there are two
ways of getting $H= U(1)$ as a subgroup, namely:
$$
\eqalign{
i) A &= 0
\cr
B &= \hat K
\cr
ii)\hat K &=\gamma A
\cr
B &=A}
\eqno (3.17)
$$
(where $B$ is the remaining abelian gauge field). The case
 (3.17{\it i}) carries the
following representation of $SU(3)\times SU(2)\times U(1)_B$ (in terms of a
real
parameter $x\neq 0$):
$$ (1,1,0) \oplus (1,2,0) \oplus (\bar 3,1, x)
\oplus (\bar 3,1,x)\oplus (3,2,-x)
\eqno (3.18)
$$
Now it is plain that this is anomalous, because, for example,
 $ Tr Y_A \{ T^{a}_{SU(2)},T^{b}_{SU(2)} \} = -6x \neq 0 $

On the other hand, among the representations given by (3.17{\it ii}),
 and parametrized
by $\gamma$, there is a unique one which enjoys the property of being anomaly
free, because then the $U(1)$ fermion quantum numbers are
$$2x\oplus -x \oplus(1+\gamma)x \oplus(\gamma -1)x\oplus -\gamma x
\eqno (3.19)
$$
so that the condition of vanishing of $Tr Y_{B}^3 = 6 + 18\gamma$ uniquely
fixes $\gamma = -1/3$. This yields again the anomaly free representation
given before in eq. (3.1). If we substitute $\gamma = -1/3$ in eq.
(3.19{\it ii}), one obtains $ A= -tr K$, namely the unimodularity condition.
\par
We shall now,
for the sake of completeness, generalize the preceding discussion to the case
in which there exists a right handed neutrino.
In the old NCG scheme the fermions (namely, now $\bar e_L,\bar \nu_L,l_L,
\bar d_L,\bar u_L,q_L$) carry the following representation of $G$:
$$
(1,1,y,y^{\prime},0)\oplus(1,1,\!-y,y^{\prime},0)\oplus(1,2,0,\!-y^{\prime},0)
\oplus(\bar3,1,y,0,k)\oplus (\bar 3,1,\!-y,0,-k)\oplus(3,2,0,0,-k)
\eqno (3.20)
$$
It is readily seen that this representation is anomalous, because
$ Tr Y_{A^{\prime}} \{ T^{a}_{SU(2)},T^{b}_{SU(2)} \} = -2 y^{\prime}
\neq 0$.
We next look for subgroups of $G$ having the $SU(3)\times SU(2)$
structure given by eq. (3.2). There are again only two possibilities,
either $H=U(1)\times U(1)$ or else $H=U(1)$. The representations of
$U(1)\times U(1)$ induced by (3.20) are obtained by imposing on the abelian
gauge fields $A$, $A^{\prime}$ and $\hat K$, the linear restrictions spelled
out
in (3.6). It is not difficult to see that the subcases (3.6{\it ii}) and
always lead to anomalous representations. This result is indeed the same as in
the massless neutrino situation.
However, at variance with this case, there is now an anomaly free
representation
in the subcase (3.6{\it i}). Actually, (3.6{\it i}) leads to the
     following representations
of $H= U(1)_B\times U(1)_{B^{\prime}}$:
$$((1+\alpha)x,\beta y)\oplus ((\alpha-1)x,\beta y)\oplus (-\alpha x,-\beta y)
\oplus (x,y)\oplus(-x,y)\oplus(0,-y)
\eqno (3.21)
$$
Absence of anomalies implies $Tr Y_B^3 = 6\alpha x^3 =0 $ and
$ Tr Y_{B^{\prime}}Y_{B}^2
= (2\beta+6)y x^2 =0$, which is only possible if $\alpha =0$ and $\beta =-3$.
(Recall that both $x,y \neq 0$, because they set the charge scale).
Now, if one substitutes this in (2.23), and works out the remaining anomaly
constraints (including mixed gauge-gravitational anomalies), one easily checks
that they all hold.
Besides, the hypercharge assignments are again exactly the same as the ones
obtained using unimodularity. Plugging the values for $\alpha$ and $\beta$
back in (3.6) we obtain the constraint $A^{\prime} = -tr K$, while $A$ remains
free. (All this is the same as unimodularity [\Proteus ]).
Arguing from the standpoint of absence of anomalies,
we have shown that, within the old scheme, we need at least a family with no
right-handed neutrino, if reduction to the standard model representation for
$SU(3)\times SU(2) \times U(1)$ is to be achieved.
Exactly the same conclusion can be reached from the standpoint of
unimodularity [\Proteus].
\par
To close this section, we shall comment on the equivalence between
unimodularity
and absence of anomalies with right handed neutrinos in the framework of the
new
scheme. Performing exactly the same type of analysis as we did repeatedly, one
again reaches the conclusion that both viewpoints are equivalent, and both
 reduce
the gauge group from $SU(3)\times SU(2)\times U(1)_A\times U(1)_K$ down to the
standard model group $SU(3)\times SU(2)\times U(1)$ with the correct
representation content.

\section 4. Conclusions and comments

Non-Commutative Geometry has provided us, up to now, with a way of getting
{\it some} lagrangians in  field theory, corresponding to the
standard model, with certain relationships among the parameters; not every set
of coupling constants
can be obtained from this viewpoint [\Sirius],[\Book],[\KStris].
If one, however, now takes this {\it classical} lagrangian as the starting
point
 for a {\it quantum} theory, these relationships are not maintained (because
they are not first integrals of the renormalization
 group)[\Orpheus],[\Persephone].
In the present work, we have pointed out at a curious fact, {\it unimodularity}
, which, being as it is a mathematical restriction on the group of unitaries
of a certain algebra, has been shown to be equivalent to the
{\it physical} condition of absence of anomalies in the model.

We would like now to make some comments concerning the matter of anomaly
cancellation in the NCG framework. We should  stress that the
constraints implying
cancelation of anomalies involving gravitons have never been used.
In the NCG formulation of the Standard Model, once anomaly
freedom for the representation of the gauge group is achieved, anomaly
freedom of the theory coupled to gravity follows. This is at variance
with the ordinary derivation of the Standard Model using standard
differential geometry, where, actually, the cancelation of the triangle
anomaly involving a $U(1)$ field and two gravitons is needed to restrict
the allowed hypercharges  as much as possible [\Ibanez]. Finally, it is
remarkable that the bizarre solution to the anomaly freedon equations
found by Minaham et al. [\Ibanez] does not occur in the NCG
reconstruction of the Standard Model. The bizarre solution is given by
the following hypercharge, $Y$, assignments
$$
Y(\bar{ e}_L) = Y( l_L) = Y (q_L) = 0,\;\;
Y(\bar {d}_L) = -Y(\bar u_L)\neq 0
$$
For these assignments to occur in NCG the following linear relations
amomg the $U(1)$ fields in eq. (2.3) should hold
$$
A+A^{\prime} = 0,\;\; A^{\prime}=0,\;\; Tr K = 0,
$$
The fermions $\bar{e}_L$, $l_L$ and $q_L$ have thus vanishing
hypercharges. But then the preceding linear relations imply
$$
A+Tr K = 0,\;\; -A+Tr K = 0,
$$
so that $\bar{d}_L$ and $\bar{u}_L$ have unavoidably zero hypercharge.
The bizarre solution thus evaporates.
A similar argument can be devised to explain the lack of bizarre
solution in the new scheme.
We would like to finish this article by saying that the results
presented in it hint at a  deeper relationship between quantum
physics
and NCG than was thought before. It certainly remains a fascinating avenue
to explore further.

\bigskip\bigskip

\leftline{\bf Acknowledgements}
We thank A. Connes, M.B. Gavela, L.E. Ib\'a\~nez, S. Peris and J. C. V\'arilly
for useful
 discussions.
This work has been supported in part by CICYT (Spain).
\bigskip\bigskip

\leftline{\bf References}
\frenchspacing
\vglue 4truemm

\refno\Sirius. J. C. V\'arilly and J. M. Gracia-Bond{\'\i}a, J. Geom.
Phys. 12 (1993) 223.

\refno\Book. A. Connes, Non-Commutative Geometry (Academic Press,
London, 1994).

\refno\KStris. D. Kastler and T. Sch\"ucker, preprint hep-th/9501077,
CPT, Luminy, 1995.

\refno\Tresttalk. A. Connes, talks given at the Conference on Noncommutative
Geometry and its Applications, T\v re\v s\v t, Czech Republic, May 1995.

\refno\Proteus. J. M. Gracia-Bond{\'\i}a, ``Connes' interpretation of
the Standard Model and massive neutrinos", to appear in Phys. Lett. B.

\refno\Connesprivatecomm. A. Connes, private communication.
\refno\lag. L. Alvarez-Gaum\'e, Anomalies in Quantum Field Theory
(Carg\`ese
 lectures)

\refno\Ibanez. L.E.Ib\'a\~nez, Proceedings of the 5th ASI on Techniques
and Concepts of High Energy Physics, St.Croix (Virgin Islands), July
14-25 1988 Edited by T. Ferbel (Plenum Press, 1989);
C. Geng and R. Marshak,
Phys. Rev, D39 (1989) 693;
A. Font et al, Phys. Lett.,B228 (1989) 79;
J. Minahan et al, Phys. Rev, D41 (1990) 715;
K. Babu and R. Mohapatra, Phys. Rev, D41 (1990) 271.
\refno\Orpheus. E. Alvarez, J. M. Gracia-Bond{\'\i}a and C. P.
Mart{\'\i}n, Phys. Lett. B306 (1993) 55.

\refno\Persephone. E. Alvarez, J. M. Gracia-Bond{\'\i}a and C. P.
Mart{\'\i}n, Phys. Lett. B329 (1994) 259.

\bye